\documentclass[a4paper,11pt]{article}

\usepackage[a4paper,margin=1in]{geometry}

\usepackage[numbers,compress]{natbib}
\usepackage{bibunits}
\defaultbibliography{549}
\defaultbibliographystyle{vancouver}

\usepackage{graphicx}
\usepackage[singlespacing]{setspace}

\usepackage[utf8]{inputenc} 
\usepackage[T1]{fontenc}    
\usepackage{url}            
\usepackage{booktabs}       
\usepackage{amsfonts}       
\usepackage{amsmath}        
\usepackage{scrtime}
\usepackage{booktabs}
\usepackage{multirow}
\usepackage{titling}
\usepackage{url}

\newcommand{\C}{\mathcal{C}}
\newcommand{\V}{\mathcal{V}}
\newcommand{\U}{\mathcal{U}}

\newcommand{\br}{\mathbf{r}}

\newcommand{\hl}{\textsuperscript{\tiny *}}

\title{A model for structured information representation in neural networks}

\author{ Michael G.~M\"uller$^{1}$, Christos
H.~Papadimitriou$^2$, Wolfgang Maass$^1$,\\ Robert Legenstein$^{1}$\\ $^1$ Institute of Theoretical
Computer Science, Graz University of Technology,\\ Graz, Austria,
\texttt{\{mueller,legi,maass\}@igi.tugraz.at}.\\ $^2$ EECS, UC Berkeley, CA
94720, USA, \texttt{christos@cs.berkeley.edu} }

\newcommand{\authorshort}{ Michael G.~M\"uller,
Christos H.~Papadimitriou, Wolfgang Maass,\\  Robert Legenstein  }

\makeatletter
\let\titleref\@title

\begin{document}

\maketitle

\vspace{0.5cm}

\begin{abstract}

Humans possess the capability to reason at an abstract level and to structure
information into abstract categories, but the underlying neural processes have
remained unknown.  Experimental evidence has recently emerged for the
organization of an important aspect of abstract reasoning: for assigning words
to semantic roles in a sentence, such as agent (or subject) and patient (or
object). 
Using minimal assumptions, we show how such a binding
of words to semantic roles emerges in a generic spiking neural network
through Hebbian plasticity.  The resulting model is consistent
with the experimental data and enables new computational functionalities such
as structured information retrieval, copying data, and comparisons. It thus
provides a basis for the implementation of more demanding cognitive
computations by networks of spiking neurons.
\end{abstract}

\bigbreak


\textbf{Running title:} Structured representations in neural networks

\bigbreak

\begin{bibunit}
\section{Introduction}

Consider the two sentences ``The truck hit the ball'' and ``The ball
hit the truck''. Even though they both consist of the same set of
words, they convey fundamentally different information which is easily
distinguished by a human listener. In order to do so, sentence
information is represented in a structured manner in cortex that takes
semantic roles of words (e.g., agent or patient) into account \citep{FranklandGreene:15}.

We can describe this problem on three different levels, corresponding
to Marr’s levels of analysis \citep{marr1982vision}. The computational
level describes the high-level computational goal, which is the generation of a structured sentence
representation.  On the second, algorithmic level, we ask how
structured representations can be implemented algorithmically. Recent experimental evidence (see below)
suggests that on this algorithmic level,
one predefined variable exist for each semantic
role (we term them semantic variables). During language processing, contents (words) are
assigned to these variables accordingly, an operation  that is referred to as binding \citep{plate1995holographic,van2006neural,eliasmith2012large,ajjanagadde1989efficient,shastri1999advances,hayworth2012dynamically,kriete2013indirection,zylberberg2013neuronal,hayworth2018thalamic,FriedericiSinger:15}.  Finally, on the
implementation level, these operations are implemented by the neuronal
networks of the brain. In this work, we establish a connection between
the latter two levels by showing how structured representations can
emerge in a spiking neural network based on well-known
neurophysiological mechanisms, and with minimal further assumptions.

A number of recent experimental findings shed some light onto the question of
stimulus representations in the medial temporal lobe (MTL). It has
been shown that ``concept cells'' located in the MTL underlie the
representation of stimulus identity \citep{quiroga2016neuronal}. Each
concept cell responds to some unique concept present in the stimulus
regardless of the mode of presentation, thus giving rise to a sparse
representation of inputs. It has furthermore been shown that
assemblies of concept cells in the MTL are not static, but can rapidly
form new associations \citep{IsonETAL:15}.

In another key experiment,
Frankland and Greene \citep{FranklandGreene:15} investigated the
cortical response to simple sentences like those discussed above using
functional magnetic resonance imaging (fMRI; see also
\citep{wang2016identifying}).  They showed that for some semantic
variables (such as the agent or patient in a sentence), there exist
specific subareas of the left mid-superior temporal cortex (lmSTC)
which encode these variables.  Importantly, after presenting some
sentence, the content of these variables could be read out from the
corresponding subareas, indicating that these variable-specific areas
form content specific representations in the binding process. This
finding, which cannot be reproduced by most existing binding models
(see {\em Discussion}) forms the basis for the model presented in this
article.

In the general form of binding, one or multiple features are bound to an
object; this problem has been addressed by a host of models (see e.g.,
\citep{plate1995holographic,van2006neural,eliasmith2012large,FriedericiSinger:15,von1994correlation}).
In the specific variant of the binding problem considered here, a semantic variable is bound to a word in a sentence (one could describe this more generally as an abstract category being bound to some content, but we stay in the following in the realm of language processing).  Models for binding
typically rely on specifically constructed circuitry or specific connectivity
between individual neurons or groups of neurons. It is unclear whether such detailed assumptions can be
justified using the available data on cortical connectivity. A further open
question is whether the specific wiring inherent to these models can emerge
through plasticity mechanisms known to shape cortical networks.

To model the emergence of structured representations in view of the above experimental findings, we
take a different approach: using minimal assumptions, we
study a generic spiking neural
network model solely using mechanisms supported by a large number of experimental
data. In particular, we make no
assumptions on specific wiring or symmetric connectivity.
We model each variable-encoding subarea of the lmSTC  -- as identified by Frankland and Greene \citep{FranklandGreene:15} -- as one population of
neurons with divisive inhibition \citep{carandini2012normalization,wilson2012division,legenstein2017probabilistic,jonke2017feedback}. The full model may have several such
``variable spaces'', one for each represented semantic variable. These variable spaces
are sparsely connected to a single content space, where neuronal assemblies
represent content which could be a word or a value. Hence, neurons in the
content space model concept cells in the MTL \citep{quiroga2016neuronal}.
We propose that the
control over binding processes between these spaces is implemented through disinhibition
\citep{Pfeffer:14,HarrisShepherd:15,LetzkusETAL:15,Caroni:15,FuETAL:15,FroemkeSchreiner:15}.
When a variable space is disinhibited while an assembly is active in content space, a sparse assembly also emerges there.
Fast Hebbian plasticity ensures that this assembly is stabilized and synaptically linked to the content space
assembly.  In other words, the variable space now contains a projection of the
content space assembly. We refer to such a projection of an assembly to a
variable space as an assembly projection. We propose that information about this assembly is retained in the variable space through transiently increased neuronal excitabilities \citep{LetzkusETAL:15}.

We show that this model can represent simple sentences in a structured manner
by binding content from the content space to semantic variables encoded by the variable
spaces. The resulting activity of the network is consistent with the fMRI data
from Frankland and Greene \citep{FranklandGreene:15}. We further show that this model is capable of
performing additional elementary processing operations proposed as atomic
computational primitives in the brain by Marcus et al.~\citep{MarcusETAL:14}:
copying the content of one variable to another, and comparing the contents of
two variables. Our results indicate that the concept of neural spaces and
assembly projections extends the computational capabilities of neural networks
in the direction of cognitive computation, symbolic computation, and the
representation and computational use of abstract knowledge.

The presented model combines and extends ideas of previously proposed
pointer-based and anatomical binding models
(e.g.~\citep{zylberberg2013neuronal,hayworth2018thalamic}, see {\em
  Discussion} for a comparison to existing models). By combining these
ideas, we show the following key advances: (1) we show that binding of
semantic variables to content is possible without the need for
specific circuitry and wiring patterns, rather, binding emerges in
generic spiking networks under Hebbian plasticity even under sparse network
connectivity, (2) the encoding of information in the network is
consistent with recent findings in the MTL (assembly representations
through concept cells \citep{quiroga2016neuronal}), (3) the model is
consistent with recent fMRI data on structured sentence encoding in
lmSTC \citep{FranklandGreene:15}, (4) the same model is capable of
performing additional elementary processing operations
\citep{MarcusETAL:14}, and (5) we prove the operativeness of the
assembly pointer principle in a detailed spiking neural network model
based on experimentally well-supported mechanisms.

\section{Results}

\subsection{Generic network model for assembly projections}

Experimental data obtained by simultaneous recordings from large sets of
neurons through multi-electrode arrays or Ca$^{2+}$ imaging showed that neural
activity patterns in cortex can be characterized in first approximation as
spontaneous and stimulus-evoked switching between the activations of different
(but somewhat overlapping) subsets of neurons (see e.g.
\citep{Buzsaki:10,BathellierETAL:12,LuczakMacLean:12}). These subsets of
neurons are often referred to as neuronal assemblies.
Consistent with this experimental data, information about specific
words/concepts and variables is represented in our model by corresponding
assemblies of neurons in generic neural circuits, referred to as neural spaces
in the following. In particular, we assume that specific content -- which could
be words, concepts, values, etc.~--  are encoded by assemblies in a neural
space $\C$, which we refer to as the {\em content space}, see
Fig.~\ref{fig:illustration}A.

Frankland and Greene showed that for some semantic variables (such the agent or
patient in a sentence), there exist specific subareas of the lmSTC which encode
these variables \citep{FranklandGreene:15}.  We model such subareas for
variables $u, v, \dots$ as a distinct set of neural spaces $\U, \V, \dots$ and
refer to these spaces as {\em variable spaces} in the following.  Each variable
space can be viewed as functioning like a register in a computer
\citep{FranklandGreene:15}. But in contrast to such registers, variable spaces
do not store content directly.  Instead, we hypothesize that a storage
operation leads to the emergence of an assembly in the variable space, in
particular an assembly which is linked via strong synapses to the assembly
representing the particular content in the content space. We call these
projections of assemblies to variable spaces {\em assembly projections}.  In
other words, by assembly projection we mean the emergence of an assembly in a
variable space as the result of afferent activity by an assembly in the content
space, with the establishment of strong synaptic links from the content
assembly to the new one.  From a computational viewpoint, such an assembly
projection is a ``pointer'' (or ``handle'') to a content in the content space.

Importantly, we do not assume specifically designed neural circuits which
enable the creation of such assembly projections and thus variable binding.
Instead, we assume a rather generic network for each variable space and the
content space, with lateral excitatory connections and lateral inhibition
within the space (a common cortical motif, investigated e.g.~in
\citep{jonke2017feedback}). Furthermore, we assume that neurons in the content
space are sparsely connected to neurons in variable spaces and vice versa, see
Fig.~\ref{fig:illustration}A. We will show that the creation of an assembly
projection in a variable space -- implementing the binding of a variable to
some content -- emerges naturally in such generic circuits with random
connectivity from plasticity processes.

In addition, our model takes into account that neurons typically do not fire
just because they receive sufficiently strong excitatory input. Experimental
data suggest that neurons are typically prevented from firing by an
``inhibitory lock'' which balances or even dominates excitatory input
\citep{HaiderETAL:13}. Thus, a generic pyramidal cell is likely to fire because
two events take place: its inhibitory lock is temporarily lifted
(``disinhibition'') and its excitatory input is sufficiently strong.
A special type of inhibitory neuron (VIP cells) has been identified as a likely
candidate for triggering disinhibition, since VIP cells target primarily other
types of inhibitory neurons (PV+ and SOM+ positive cells) that inhibit
pyramidal cells (see e.g. \citep{HarrisShepherd:15}).  Firing of VIP cells is
apparently often caused by top-down inputs (VIP cells are especially frequent
in layer 1, where top-down and lateral distal inputs arrive). Their activation
is conjectured to enable neural firing and plasticity within specific patches
of the brain through disinhibition (see e.g.
\citep{LetzkusETAL:15,Caroni:15,Pfeffer:14,FuETAL:15,FroemkeSchreiner:15}). One
recent study also demonstrated that long-term plasticity in the human brain can
be enhanced through disinhibition \citep{CashETAL:16}. We propose that top-down
disinhibitory control plays a central role for neural computation and learning
in cortex by initiating for example the creation and reactivation of assembly
projections. We note that we are not investigating the important question:
which neural processes resulted in the decision to disinhibit this particular
variable space – that is, to decide whether a word is the agent or patient of a
sentence.
We modeled disinhibition of neural spaces in the following way. As a default,
excitatory neurons received an additional strong inhibitory current that
silenced the neural space. The space was disinhibited by the removal of this
inhibitory current, which enabled activity in the neural space if excitatory
input was sufficiently strong (see {\em Supplement S1}).

\subsection{Emergence of assembly projections through Hebbian plasticity}

\begin{figure*}[t]
\begin{center}
\includegraphics[width=\textwidth]{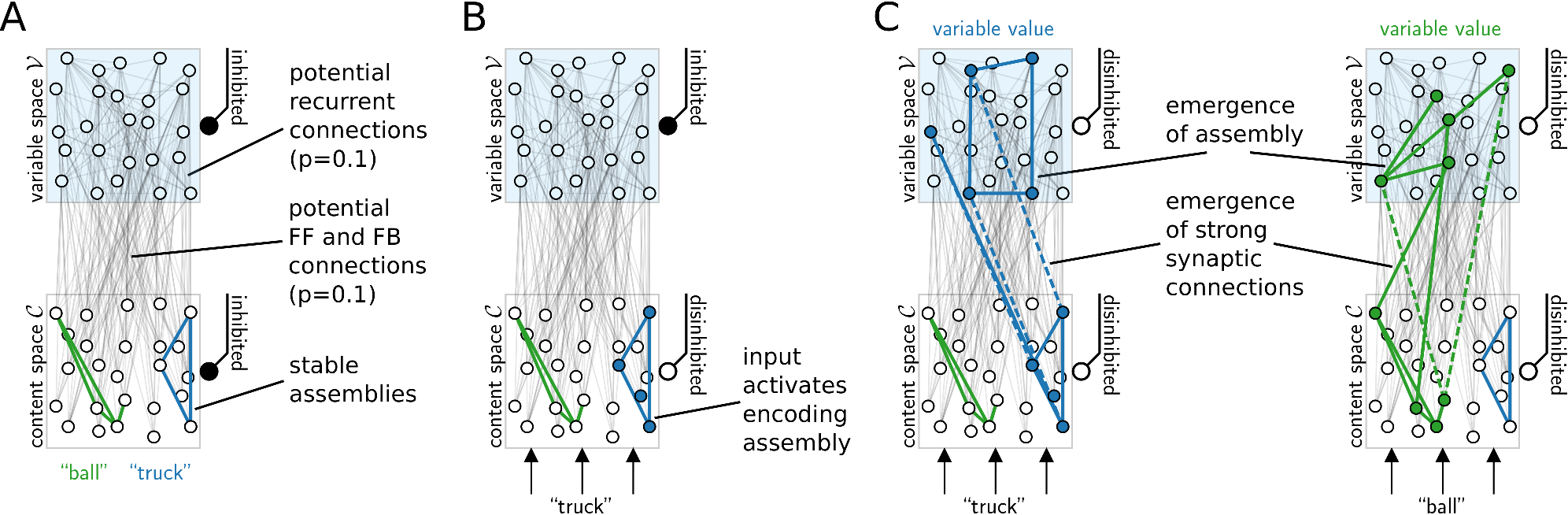}
\end{center}
\caption{
{\bf Neural spaces and assembly projections.} {\bf A)} Network
structure. Rectangles indicate content space $\C$ (white, bottom) and variable
space $\V$ for variable $v$ (light blue, top). Circles denote neurons (open:
inactive; filled: active). Disinhibition is shown on the right of each space
(filled black circle: inhibited; open circle: disinhibited). Concepts are
encoded in the content space through stable assemblies (only two assemblies for
two contents ``ball'' and ``truck'' are shown, strong recurrent connections are
depicted with colors based on assembly identity). Thin gray lines between $\C$
and $\V$ as well as within $\V$ show potential feedforward, feedback, and
recurrent connections (randomly initiated, $p = 0.1$). Only a subset of 25
neurons is shown per space.  {\bf B)} Presentation of a concept (bottom,
arrows) activates the encoding assembly in the content space if it is
disinhibited (filled dots: active neurons). Initially, the variable space $\V$
is inhibited, preventing plasticity from occurring at the potential
feedforward/feedback and recurrent connections.  {\bf C)} Disinhibition of the
variable space enables the activation of neurons there and results in the
formation of an assembly projection. Hebbian plasticity at the potential connections leads to
the emergence of an assembly in the variable space (strong recurrent
connections are shown in $\V$) which has strong synaptic connectivity with the
active assembly in $\C$ (solid lines: feedforward connections, i.e. from $\C$
to $\V$; dashed: feedback). This mechanism allows the assignment of values to
the variable encoded by the variable space. Assignment of different values
(left: ``truck'', right: ``ball'') give rise to different assembly projections
(i.e. different assemblies in $\V$ and different connectivity between $\V$ and
$\C$, color encodes identity of the value assigned to the variable space).}
\label{fig:illustration}
\end{figure*}

To test whether the binding of content to variables can be performed
by generic neural circuits, we performed computer simulations where
stochastically spiking neurons were embedded in the following network
structure (see Fig.~\ref{fig:illustration}A): The network consisted of
a content space $\C$ with 1000 excitatory neurons and a single
variable space $\V$ for a variable $v$ consisting of 2000 excitatory
neurons. In both spaces, neurons were recurrently connected
(connection probability $0.1$) and lateral inhibition was implemented
by means of a distinct inhibitory population to ensure sparse
activity. Connections between $\C$ and $\V$ were introduced randomly
with a connection probability of $0.1$. Neurons in the content space
additionally received connections from $200$ input neurons whose
activity indicated the presence of a particular input
stimulus. Hebbian-type plasticity is well-known to stabilize
assemblies (see e.g.~\cite{litwin2014formation,pokorny2017associations}) and it can also strengthen connections
bidirectionally between variable spaces and the content space. In our
model, this plasticity has to be reasonably fast so that connections
can be strengthened within relatively short stimulus presentations. As
a Hebbian-type plasticity, we used in our spiking neural network model
spike timing-dependent plasticity (STDP; \citep{bi1998synaptic,caporale2008spike}) for all synapses between
excitatory neurons in the circuit (see {\em Supplement S1} for more
details). Other synapses were not plastic.

\paragraph{Emergence of content assemblies} We do not model the initial
processing of speech, which is in itself a complicated process. Instead, we
assume that assemblies in content space act as tokens for frequently observed
input patterns that have already been extracted from the sensory stimulus.
Hence, we induced assemblies in content space $\C$ by an initial repeated
presentation of simple rate patterns provided by $200$ spiking input neurons
(see {\em Supplement S2}). We first defined five such patterns $P_1, \dots,
P_5$ that modeled the input to this space when a given content (e.g., the word
``truck'' or ``ball'') is experienced. These patterns were repeatedly presented
as input to the disinhibited content space (the value space remained inhibited
during this presentation). Due to these pattern presentations, an assembly
$\C(P_i)$ emerged in content space for each of the patterns $P_i$ (assembly
sizes typically between $50$ and $90$ neurons) that was robustly activated
(average firing activity of assembly neurons $>50$ Hz) whenever the
corresponding pattern was presented as input, see Fig.~\ref{fig:illustration}B.
STDP of recurrent connections led to a strengthening of synapses within each
assembly, while synapses between assemblies remained weak (see {\em Supplement
S2} for details).

\paragraph{Emergence of assembly projections} Assume that an active assembly
$\C(P)$ in content space represents some content $P$ (such as the word
``truck'').  A central hypothesis of this article is that disinhibition of a
variable space $\V$ leads to the creation of an assembly projection $\V(P)$ in
$\V$. This projection $\V(P)$ is itself an assembly (like the assemblies in the
content space) and interlinked with $\C(P)$ through strengthened synaptic
connections.

To test this hypothesis, we next simulated disinhibition of the variable space
$\V$ while input to content space $\C$ excited an assembly there. This
disinhibition of the variable space allowed spiking activity of some of the
neurons in it, especially those that received sufficiently strong excitatory
input from a currently active assembly in the content space. STDP at the
synapses that connected the content space $\C$ and the variable space $\V$ led
to the stable emergence of an assembly $\V(P_i)$ in the variable space within
one second when some content $P_i$ was represented in $\C$ during disinhibition
of $\V$, see Fig.~\ref{fig:illustration}C. Further, plasticity at recurrent
synapses in the variable space $\V$ induced strengthening of recurrent
connections within assemblies there (see Fig.~\ref{fig:illustration}C and {\em
Supplement S2}). Hence, disinhibition led to the rapid and stable creation of
an assembly in the variable space $\V$, i.e., an assembly projection. We denote
such creation of an assembly projection in a variable space $\V$ for a specific
variable $v$ to content $P$ encoded in content space by $\text{CREATE}(v,P)$.

\begin{figure*}[t!]
\begin{center}
\includegraphics[width=\textwidth]{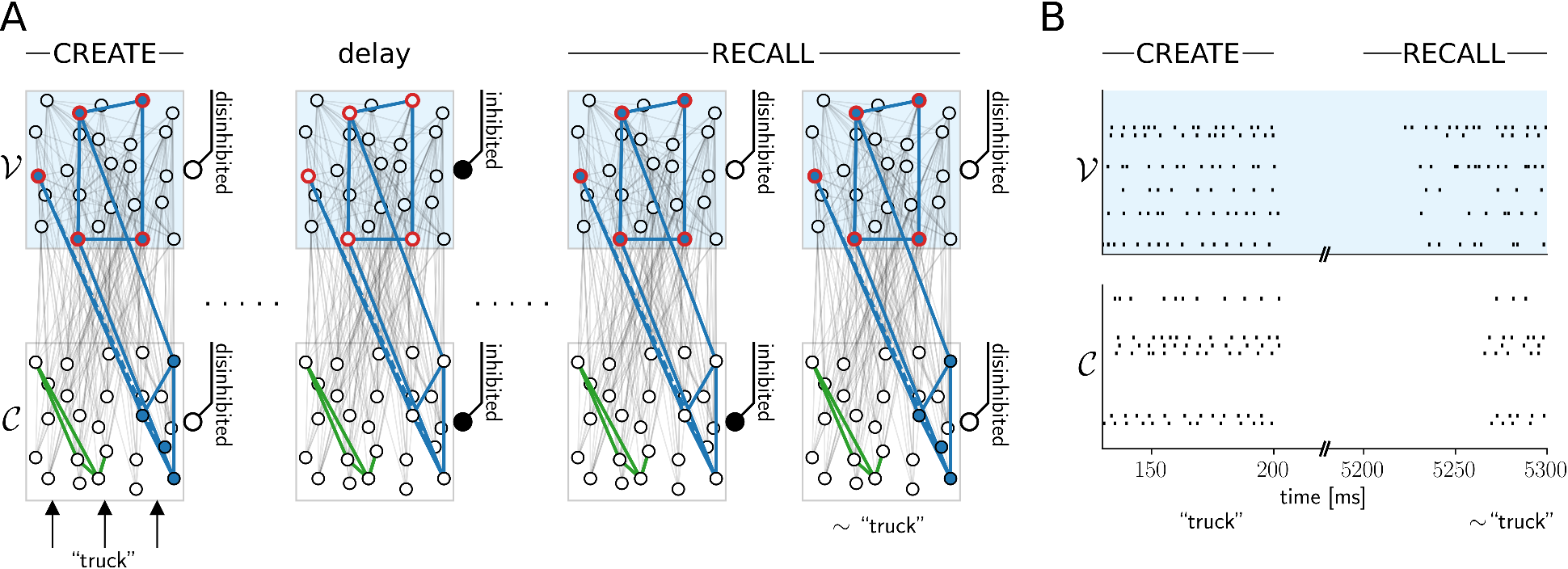}
\end{center}
\caption{
{\bf Variable-specific recall through assembly projections.} {\bf
A)} Recall of a previously created assembly projection (schematic drawing). After an assembly projection was
formed (CREATE) for the word ``truck'', the excitability of assembly neurons in
the variable space $\V$ is enhanced (indicated by red color). When the variable
space is disinhibited, these neurons are activated, and in turn they activate
the ``truck'' assembly in content space $\C$ (RECALL).  {\bf B)} Spike rasters
from variable space $\V$ (top) and content space $\C$ (bottom) in a simulated
recall (only $20$ randomly selected neurons shown per neural space for
clarity). After a CREATE (left, up to $200$ ms), and a delay for $5$ s, a
RECALL is initiated by first disinhibiting the variable space $\V$ (at time
$t=5200$ ms) and then disinhibiting the content space $\C$ ($50$ ms later).}
\label{fig:recall}
\end{figure*}

Fast recruitment of assemblies in a variable space necessitates rapid forms of
plasticity.  We assumed that a (possibly initially transient) plasticity of
synapses occurs instantaneously, even within seconds. The assumption of rapid
plasticity of neurons and/or synapses is usually not included in neural network
models, but it is supported by a number of recent experimental data. In
particular, \citep{IsonETAL:15} shows that neurons in higher areas of the human
brain change their response to visual stimuli after few or even a single
presentation of a new stimulus where two familiar images are composed into a
single visual scene.

\paragraph{Variable-specific recall through assembly projections} From a functional
perspective, the binding of a content to a semantic variable is exploited at a
later point in time when the content of this variable is recalled for further
processing. In our model, a recall $\text{RECALL}(v)$ of the variable's $v$
content should lead to the activation of the assembly $\C(P)$ in content space
which was active at the most recent $\text{CREATE}(v,P)$ operation (e.g.,
representing the word ``truck''). The strengthened synaptic connections between
assemblies in variable space $\V$ for variable $v$ and content space $\C$ may
in fact enable such a recall. However, an additional mechanism is necessary
that reactivates the most recently active assembly in variable space $\V$. One
possible candidate mechanism is the activity-dependent change of excitability
in pyramidal cells.  It has been shown that the excitability of pyramidal cells
can be changed in a very fast but transient manner through fast depression of
GABA-ergic synapses onto  pyramidal cells \citep{KullmannETAL:12}. This effect
is potentially related to the match enhancement or match suppression effect
observed in neural recordings from monkeys, and is commonly used in neural
network models for delayed match-to-sample (DMS) tasks (see
e.g.~\citep{TartagliaETAL:15}). Using such a mechanism, a $\text{RECALL}(v)$
can be initiated by disinhibition of the variable space $\V$ while the content
space does not receive any bottom up input, see Fig.~\ref{fig:recall}A. The
increased excitability of recently activated neurons in $\V$ ensures that the
most recently active assembly is activated which in turn activates the
corresponding content through its (previously potentiated) feedback connections
to content space $\C$.

We tested whether such a mechanism can reliably recall previously bound
contents from variables in our generic network model. A transient increase in
neuron excitability has been included in the stochastic spiking neuron model
through an adaptive neuron-specific bias that increases slightly for each
postsynaptic spike and decays with a time constant of $5$ seconds (see {\em
Supplement S2}).  We used an automated optimization procedure to search for
synaptic plasticity parameters that lead to clean recalls (see {\em Supplement
S2}, all parameters were constrained to lie in biologically realistic ranges).
Fig.~\ref{fig:recall}B shows the spiking activity in our spiking neural network
model for an example recall $5$ seconds after the creation of the assembly
projection. One sees that the assembly pattern that was active at the create
operation was retrieved at the recall.

In general, we found that the contents of the projection can reliably be
recalled in our model.  In order to test whether the model can deal with the
high variability of individual randomly created networks as well as with the
variability in sizes and connectivity of assemblies, we performed several
simulations with random network initializations.  We used five different
content space instances with randomly chosen recurrent connections. In each of
those, we induced five stable assemblies encoding different contents as
described above. Note that these assemblies emerged through a plasticity
process, so their sizes were variable. For each of these five content space
instances, we performed ten simulations where the variable space was set up and
randomly connected in each of these. This procedure ensured that we did not
test the network behavior on a particular instantiation of the circuit, but
rather whether the principle works reliably for generic randomly connected
circuits.

Performing a CREATE/RECALL sequence separately for each of the five patterns,
we found that the recall was successful in each of the $250$ trials ($5$
content spaces with $5$ patterns each, and this repeated $10$ times for
different variable spaces). A recall was regarded as a success if during the
RECALL phase, at least $80$\% of the neurons which previously belonged to the
concept assembly (measured after the assembly induction in the content space)
were active (firing rate $>50$ Hz), and if the number of erroneously active
neurons did not exceed $20$\% of the original assembly size. Typically, one or
two neurons from the original assembly were missing, but no excess neurons
fired during the RECALL.

\subsection{Reproducing experimental data on the binding of words to roles and
structured information representation}

\begin{figure*}[tp]
\begin{center}
\includegraphics[width=\textwidth]{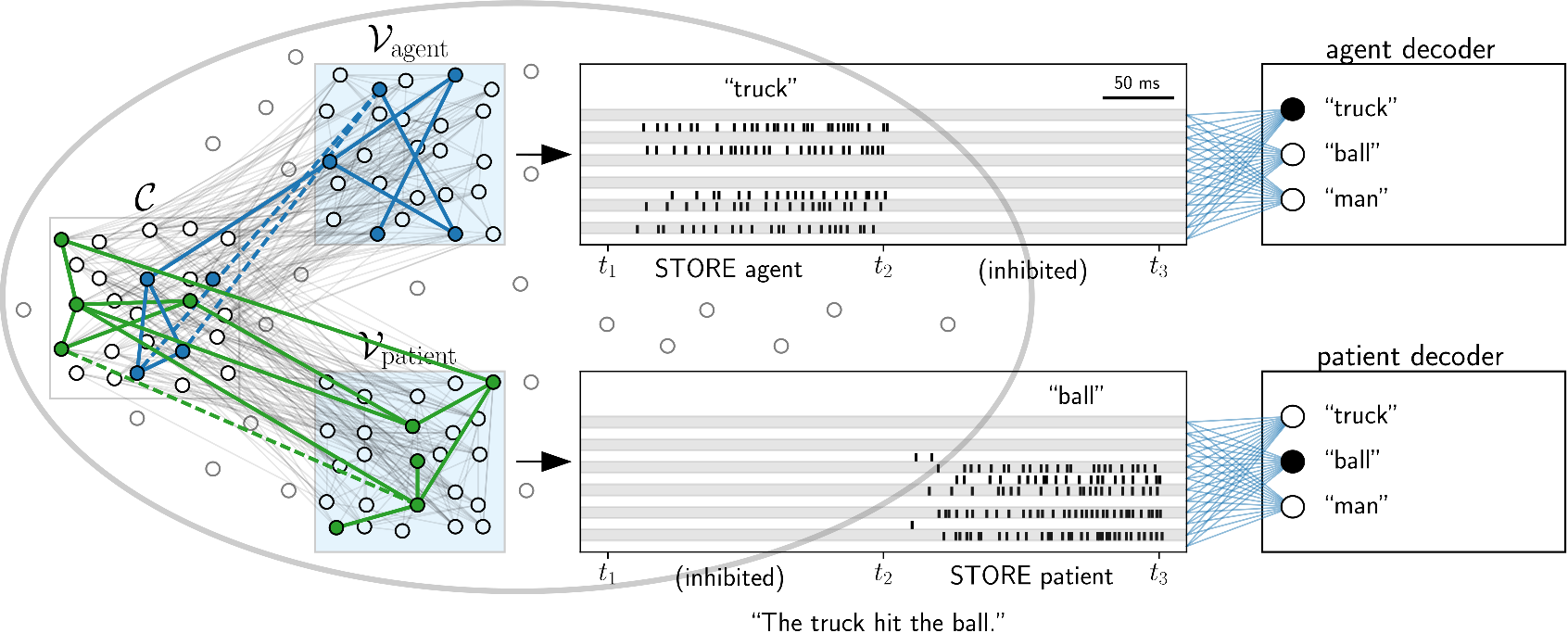}
\end{center}
\caption{
{\bf Decoding of agent/patient identity from assembly
projection, as in \citep{FranklandGreene:15}.} The content space $\C$ (white
background, left) was connected to two neural spaces for storing the agent
(light blue background, top) and the patient (light blue background, bottom) of
a sentence.  Presentation of a word lead to the activation of the corresponding
assembly in $\C$ (only the words for the agent and the patient were processed,
verbs and articles were ignored).  Words were presented in the order in which
they appear in the sentence (for ``The truck hit the ball.'', the input
``truck'' was presented while the agent space was disinhibited between $t_1$
and $t_2$, likewise disinhibition of the patient space while ``ball'' was
presented between $t_3$ and $t_4$).  This resulted in the establishment of
assembly projections within the variable spaces for agent and patient (insets
show spike rasters of a subset of neurons from each variable space).  Linear
classifiers (``agent decoder'' and ``patient decoder'') were able to determine
the current agent and patient in sentences from the corresponding variable
spaces, modeling experimental results in \citep{FranklandGreene:15}.  The
information could be decoded from the variable space.}
\label{fig:fg}
\end{figure*}

Two experiments performed in \citep{FranklandGreene:15} provided new insights
in how variables may be encoded in neocortex.  Sentences were shown to
participants where individual words (like ``truck'' or ``ball'') occur as the
agent or as the patient. The authors then studied how cortex represents the
information contained in a sentence. In a first experiment, the authors aimed
to identify cortical regions that encode sentence meaning.
Example sentences with the words ''truck'' and ``ball'' are ``The truck hit the
ball'' and ``The ball hit the truck''. The different semantics of these
sentences can be distinguished for example by answering the question ``What was
the role of the truck?'' (with the answer ``agent'' or ``patient''). Indeed,
the authors showed that a linear classifier is able to distinguish these
sentences from the fMRI signal of left mid-superior temporal cortex (lmSTC).
Using our model for assembly projections, we can model such situations by
binding words either to an agent variable (``who did it'') or to a patient
variable (``to whom it was done''). Under the assumption that lmSTC hosts
variable spaces (with assembly projections) for the role of words, it is
expected that semantic decoding is possible from the activity there, but not
from the activity in content space where the identities are encoded
independently of their role. We performed simulations where the words ``truck''
and ``ball'' (represented by some assemblies in content space) were
sequentially bound (the temporal sequence was mimicking the position of the
word in the sentence) either to variable space $\V_\text{agent}$ or
$\V_\text{patient}$, depending on their role. Note that we did not model the
processing of the verb or other words in the sentence, as only the
representation of the agent and the patient was investigated in
\citep{FranklandGreene:15}. Low-pass filtered network activity was extracted
for each of the resulting four sequences. We then trained a linear classifier
to classify the role of ``truck'' for each time point during the sequential
binding, based on a noisy version of filtered network activity. We found that
even under severe noise conditions, the classifier was able to nearly perfectly
classify test sentences (i.e., new simulations with new noisy activity on the
same sentences; test classification error $3.0$ \%). On the other hand, a
classifier based on activity of the content space performed only slightly
better than random with a test classification error of $44.4$ \%.

A second experiment in \citep{FranklandGreene:15} revealed that information
from lmSTC subregions can also be used to read out the current contents of the
variables for the agent and the patient. More specifically, the authors showed
that it is possible to predict the identity of the agent from the fMRI signal
of one subregion of lmSTC and the identity of the patient from the signal in
another subregion (generalizing over all identities of other roles and over
different verbs). We expected that this would also be the case in the proposed
model since the assemblies that are formed in the variable spaces
$\V_\text{agent}$ and $\V_\text{patient}$ are typically specific to the bound
content. We tested this hypothesis by training a multinomial logistic
regression model to classify the content of the variable for each of the two
variable spaces (agent and patient) at times when these spaces were
disinhibited (Fig.~\ref{fig:fg}, ``agent decoder'' and ``patient decoder'').
Here, we bound words to variable spaces as before, but we considered all $40$
possibilities of how $5$ items (words) $A_1, \dots, A_5$ can be sequentially
bound (for example: $A_1$ is bound first to $\V_\text{agent}$ , then $A_2$ is
bound to $\V_\text{patient}$ ; we excluded sentences where the agent and
patient is the same word). Low-pass filtered activity of a subset of neurons
was sampled at every $1$ ms to obtain the feature vectors to the classifiers
(see {\em Supplement S2}). Half of the possible sequences were used for testing
where we made sure that the two items used in a given test sentence have never
been shown in any combination in one of the sentences used for training.
Consistent with the results in \citep{FranklandGreene:15}, the classifier
achieved nearly optimal classification performance on test data (classification
error $<3$ \% for each variable space). Note that such classification would
fail if each variable space consisted of only a single assembly that is
activated for all possible fillers \citep{zylberberg2013neuronal}, since in
this case no information about the identity of the role is available in the
variable space.

\subsection{Cognitive computations with assembly
projections}\label{sec:cognitive}

\begin{figure*}[t!]
\begin{center}
\includegraphics[width=0.8\textwidth]{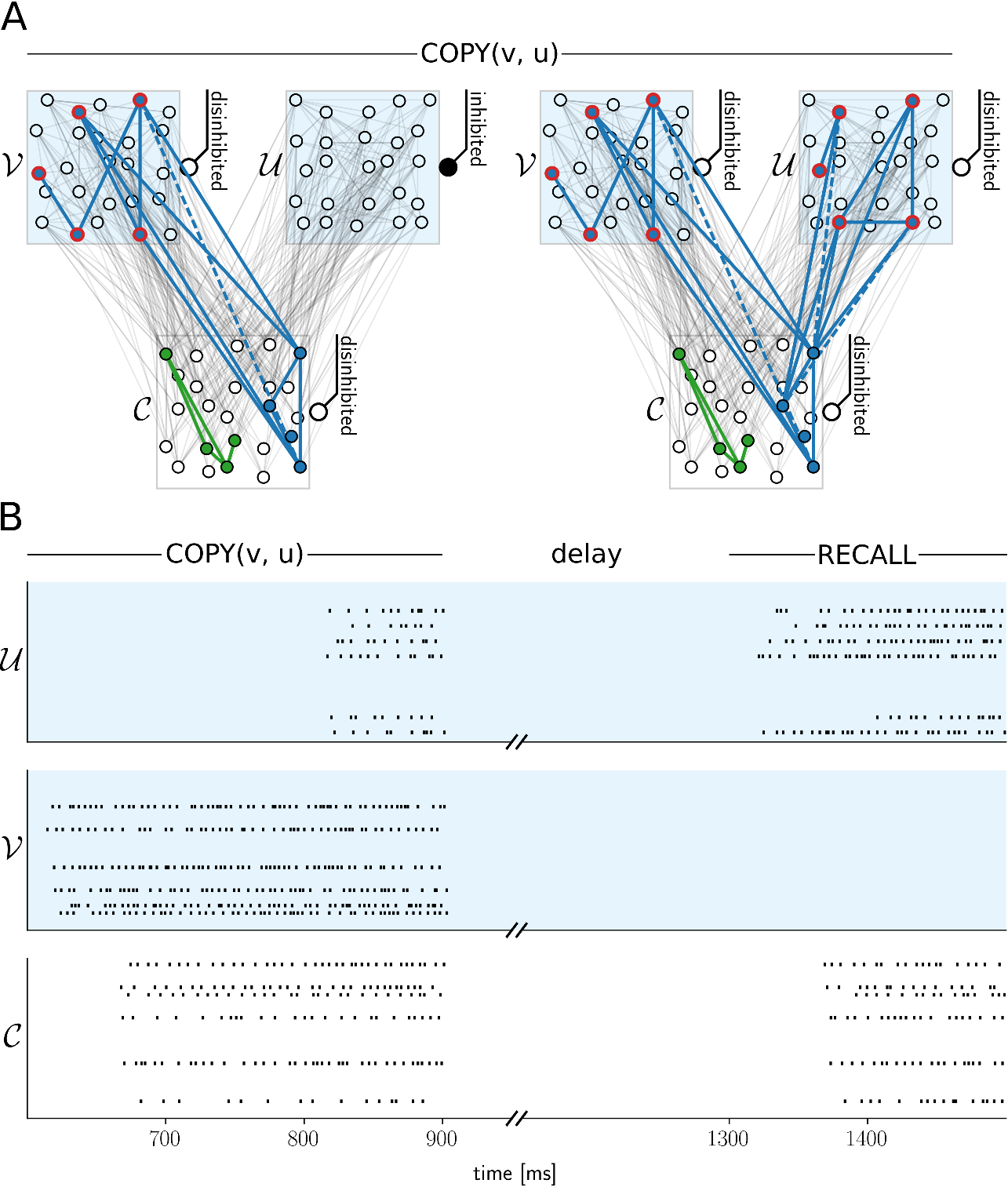}
\end{center}
\caption{
{\bf Assembly projection copy operation.} {\bf A)} Copying the content of one variable space to another variable space (schematic drawing). Disinhibition of variable space $\V$ recalls its
content in content space $\C$ (left). A subsequent disinhibition of variable
space $\U$ creates an assembly projection for this content there (right).  {\bf
B)} Spike rasters from variable spaces $\U$ (top), $\V$ (middle), and content
space $\C$ (bottom) in a simulated copy operation from a variable $v$ to a
variable $u$ ($600-900$ ms; only $20$ randomly selected neurons are shown per
neural space for clarity). After a $400$ ms delay, the content of variable $u$
is tested by a recall at time $1300$ ms. The assembly is correctly recalled
into the content space.}
\label{fig:copy}
\end{figure*}

Apart from the creation of assembly projections and recall of content, two
further operations have been postulated by Marcus et al.~to be essential for
many higher cognitive functions \citep{zylberberg2013neuronal}. The first is
$\text{COPY}(v,u)$ which copies (or routes) the content of variable $v$ to
variable $u$. In our model, the copy operation can be realized by creation of
an assembly projection in variable space $\U$ for variable $u$ to the content
to which the assembly projection in variable space $\V$ for variable $v$ refers
to. We hypothesized that this operation can be implemented in our generic
circuit model simply by disinhibiting $\V$ in order to activate the
corresponding content in $\C$ followed by a disinhibition of $\U$ in order to
create an assembly projection there, see Fig.~\ref{fig:copy}A.

To test this hypothesis, we performed simulations of our spiking neural network
model with one content space and two variable spaces. The performance was
tested through a recall from the target assembly projection $400$ ms after the
projection was copied, see Fig.~\ref{fig:copy}B. We deployed the same setup as
described above where $5$ assemblies were established in the content space,
again considering ten different pre-trained content space instances (see
above). For each of these, we performed ten copy operations (testing twice the
copying of each content assembly) and assessed the assembly active in the
content space after a recall from the target variable space. Again, all of the
100 considered cases were successful (applying the same success criterion as
before).

\begin{figure*}[t!]
\begin{center}
\includegraphics[width=0.8\textwidth]{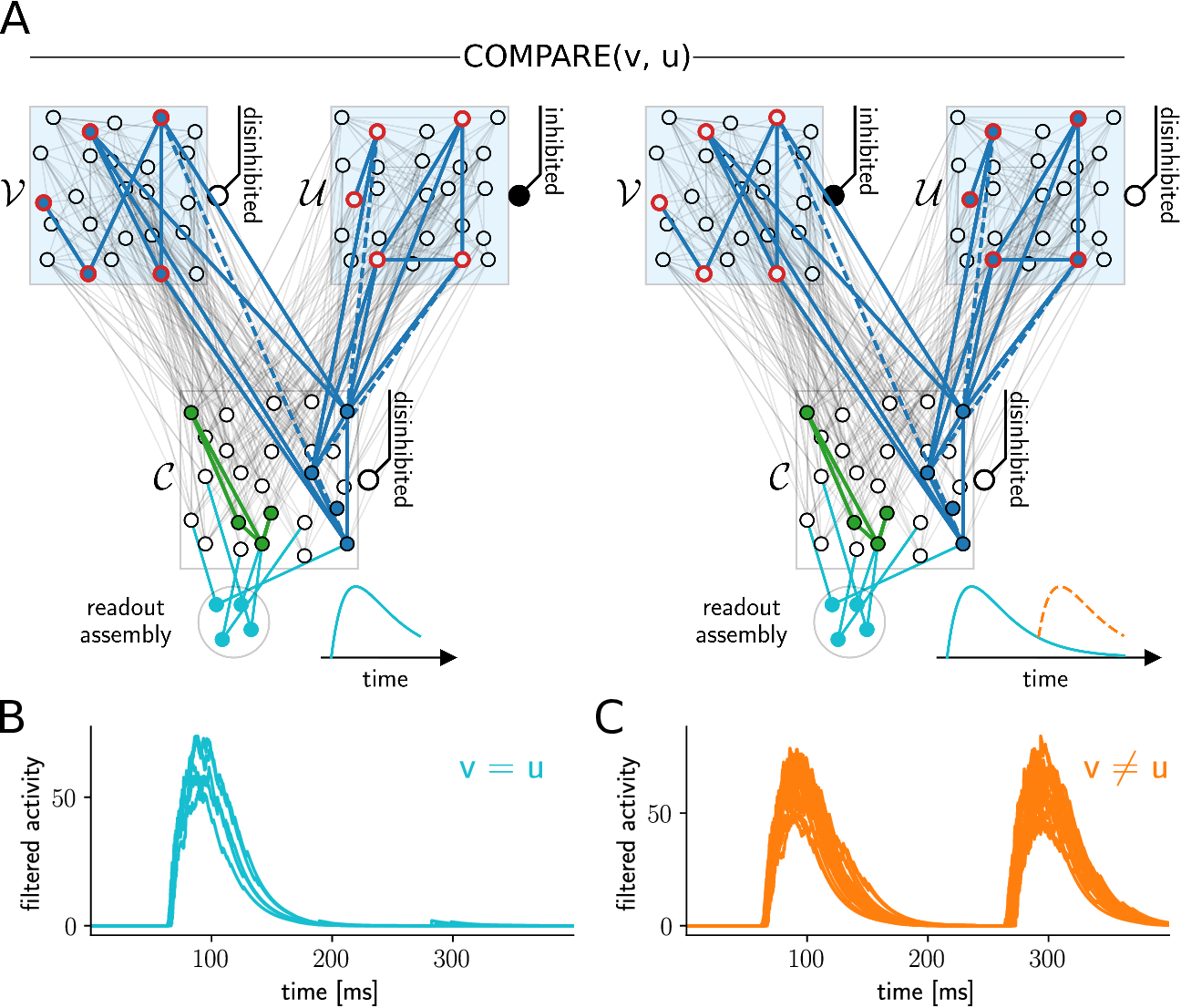}
\end{center}
\caption{
{ \bf Comparison with assembly projections.} {\bf A)}
Comparing the content of two variable spaces (schematic drawing). A population of readout neurons
(bottom, teal) receives sparse depressing connections from the excitatory
neurons in the content space. The comparison consists of a recall from variable
space $\V$ (left; red neurons indicate neurons with higher excitability)
followed by a recall from $\U$ (right). During the first recall, readout
weights become depressed and readout activity decreases (indicated by teal
trace inset right of readout). Next, a second recall is performed (right). If
the patterns are identical, the readout weights are still depressed and the
readout response is therefore weak (teal trace at readout). If the content
changes (i.e., $u \neq v$), readout weight from active neurons in $\C$ are not
depressed, which leads to strong readout activity (dashed orange trace at
readout).  {\bf B, C)} Resulting readout assembly activity in spiking neural
network model. Each trace shows the population activity (filtered with an
exponential kernel) of the readout population for one comparison operation
between two assembly projection contents ($25$ comparisons in total, one for
each possible way of assigning $5$ values to two variable spaces for variables
$v$ and $u$). At time $0$ ms, the content of variable space $\V$ was recalled
(during the first $50$ ms of each recall, the content space remains inhibited
and thus there is no readout activity) and the readout reacted in a similar
manner to all contents. From time $200$ ms on, the content of variable space
$\U$ was recalled. Due to depressed synaptic connections, the readout response
was weak when the content of $\V$ matched the content of $\U$ (panel B). In
case $\V$ and $\U$ stored different values the response was as strong as during
the first recall phase (panel C).}
\label{fig:compare}
\end{figure*}

A final fundamental operation considered in \citep{zylberberg2013neuronal} is
$\text{COMPARE}(v,u)$ which assesses whether the content of $v$ is equal to the
content of $u$. One possible implementation of this operation in our model is
established by a group of readout neurons which receive depressing synaptic
connections from the content space. Then, when the content for $\V$ and $\U$ is
recalled in sequence, the readout synapses will be depressed for the content of
$\U$ if and only if the content of $\U$ equals the content of $\V$. Such a
``change detecting'' readout population thus exhibits high activity if the
contents of $\V$ and $\U$ are different, see Fig.~\ref{fig:compare}A.
Simulation results from our spiking neural network model are shown in
Fig.~\ref{fig:compare}B, C. Using a set of $5$ contents as above, we tested
$25$ comparisons in total, one for each possibility how these $5$ contents can
be bound to two variable spaces for variables $v$ and $u$.
Fig.~\ref{fig:compare}B shows readout activity for the case when the same
content was bound to both variables $v$ and $u$ (5 cases). The readout activity
of the second recall (starting after time $200$ ms) was virtually absent in
this case. In contrast, if the two variables were bound to different content
(20 cases), the second recall always induced strong activity in the readout
population (Fig.~\ref{fig:compare}C). Hence, this simple mechanism is
sufficient to compare variables with excellent precision by simply thresholding
the activity of the readout population after the recall from the second
assembly projection.

\section{Discussion}

It has often been emphasized (see e.g.~\citep{Marcus:03,MarcusETAL:14}) that
there is a need to understand brain mechanisms for information processing via
variables.  We show in this article how binding capabilities
emerge in a generic network of spiking neurons by means of assembly
projections. Our model is consistent with recent findings on cortical
assemblies and the encoding of sentence meaning in cortex
\citep{FranklandGreene:15}.  Our neural network model is not specifically
constructed to perform such binding tasks. Instead, it is based on generic
sparsely and randomly connected neural spaces that organize their computation
based on fast plasticity mechanisms.  The model provides a direct link between
information processing on the algorithmic level of symbols and sentences and
processes on the implementation level of neurons and synapses.  The resulting
model for brain computation supports top down structuring of incoming
information, thereby laying the foundation of goal oriented ``willful''
information processing rather than just input-driven processing.
The proposed synaptic plasticity that links assemblies in different neural
spaces can be transient, but could also become more permanent if its relevance
is underlined through repetition and consolidation. This would mean that some
neurons in the variable space are no longer available to form new projection
assemblies, but this does not pose a problem if each variable space is
sufficiently large.

A large body of modeling studies have tackled the general binding problem. We
study in this work not the standard form of a binding task, where for example
one or several features are bound to an object (see, e.g.
\citep{plate1995holographic,van2006neural,eliasmith2012large,FriedericiSinger:15}
for models in that direction). Instead, we are addressing the problem of
binding abstract categories, i.e., structural information, to content. The main
classes of models in this direction are pointer-based models and models based
on indirect addressing.  Pointer-based models (e.g.
\citep{zylberberg2013neuronal}) assume that pointers are implemented by single
neurons or populations of neurons which are always co-active.  In contrast, our
model is based on the assumption that distributed assemblies of neurons are the
fundamental tokens for encoding symbols and content in the brain, and also for
projections which implement in our model some form of pointer. We propose that
these assembly projections can be created on the fly in some variable spaces
and occupy only a sparse subset of neurons in these spaces.
\citep{FranklandGreene:15} showed that the identity of a thematic role (e.g.
the agent in a sentence) can be predicted from the fMRI signal of a subregion
in temporal cortex when a person reads a sentence. As shown above, this finding
is consistent with assembly projections. It is, however, inconsistent with
models where a variable engages a population of neurons that is independent of
the bound content, such as pointer-based models. In comparison to pointer
models, the assembly projection model could also give rise to a number of
functional advantages. In a variable space $\V$ for a variable $v$, several
instantiations of the variable can coexist at the same time, since they can be
represented there by increased excitabilities of different assemblies. These
contents could be recalled as different possibilities in a structured recall
and combined in content space $\C$ with the content of other variables to in
order to answer more complex questions.

Models based on indirect addressing assume that a variable space encodes an
address to another brain region where the corresponding content is represented
\citep{kriete2013indirection}. The data of Frankland and Greene, which shows
that the stored content can be decoded from the brain area that represents its
thematic role, speak against this model in its pure form, as the address is in
general unrelated to the stored content.

Another category are anatomical binding models in which different
brain areas are regarded as distinct placeholders (similar to the
variable spaces in this work). As each placeholder may have a unique
representation for some content, anatomical binding models are
consistent with the findings of Frankland and Greene
\cite{FranklandGreene:15}. A problem in anatomical binding models lies
in providing a mechanisms which allows to translate between the
different representations. Building on previous work
\cite{hayworth2012dynamically}, a recent non-spiking model shows how
external circuitry can be used to solve this problem
\cite{hayworth2018thalamic}. Each pair of placeholders requires
multiple external processing circuits which are densely connected to
the placeholders to allow transferring and comparing contents. The
number of these processing circuits increases quadratically with the
number of placeholders. In contrast, the model presented in this work
circumvents this problem by using the content space as a central hub,
removing the need for additional circuitry.

In summary, the presented model combines the strengths of pointer-based
(e.g.~\citep{zylberberg2013neuronal}) and anatomical binding models
(e.g.~\citep{hayworth2018thalamic}).  Like anatomical binding models
\citep{hayworth2018thalamic}, the dynamics of the proposed model match
experimental data on the encoding of variables in human cortex
\citep{FranklandGreene:15}, but using the content space as a central hub
eliminates the need to add circuitry as the number of variables increases.
This is achieved by using a pointer-based mechanism, but unlike pointer models
\cite{kriete2013indirection,zylberberg2013neuronal} the proposed model is
consistent with the findings of \citep{FranklandGreene:15} and does not rely on
elaborate processing circuitry or dense bidirectional connectivity.

The validity of the assembly projection model could be tested experimentally,
since it predicts quite unique network dynamics during mental operations.
First, binding of a variable to a concept employs disinhibition of a variable
space related to that variable. This could be implemented by the activation of
inhibitory VIP cells which primarily target inhibitory neurons, or by
neuromodulatory input. Similar disinhibition mechanisms would be observed
during a recall of the filler for that variable.  Another prediction of the
model is that the assembly projection that emerges in a variable space for some
content should be similar to another one for the same content if it is
re-established on a short or medium time scale.  On the other hand, a
significant modification of the assembly that encodes a concept will also
modify the assembly projection that emerges in a variable space.  Further, our
model suggests that inactivation of an assembly projection to some content $A$
in variable space $\V$ would not abolish the capability to create a binding of
the associated variable $v$ to this content $A$: If the trial that usually
creates this binding is repeated, a new assembly projection in the variable
space for $v$ can emerge.  Finally, the model predicts that a mental task that
requires to copy (or compare) the filler of one variable $u$ to another
variable $v$ causes sequential activation (disinhibition) of the variable
spaces $\U$ and $\V$ for these variables.

The assembly projection model assumes that there is sufficient connectivity
between the content space and variable spaces in both directions such that
assemblies can be mutually excited. The most prominent language related areas
in the human brain are Broca's area and Wernicke's area. In addition, it has
been speculated that word-like elements are stored in the middle temporal
cortex \citep{berwick2016only}, corresponding to the content space in our
model.  As discussed in \citep{berwick2016only}, these areas are connected by
strong fiber pathways in adult humans. These pathways could provide the
necessary connectivity for the creation of assembly projections. The authors
further point out that some of the pathways are missing in macaque monkeys and
chimpanzees, possibly explaining the lack of human-like language capabilities
in these species.

In this paper we are presenting simulations in a realistic model that are
compatible with experimental results in \citep{FranklandGreene:15} for binding
words to roles in a sentence. Other recent experimental results
(\citep{ding2016cortical,zaccarella2015merge}, see also
\citep{friederici2017language}) seem to suggest that another assembly operation
is at work in the processing of simple sentences: the operation Merge proposed
by Chomsky as the fundamental linguistic operator mitigating the construction
of sentences and phrases \citep{chomsky2014minimalist,berwick2016only}. In
related work \citep{papadimitriou2019random}, it is shown that a MERGE
operation on assemblies can indeed be realized by neurons in a simplified
model. To illustrate MERGE, suppose that there is a third variable space where
verbs are projected, and an assembly for the word ``hit'' in content space has
been projected to this space. MERGE of the projected assemblies ``hit'' and
“truck” in variable space would then create a new assembly in another
subregion, which can be called phrase space (a brain area suspected to
correspond to phrase space is the pars opercularis of Broca’s area, or BA 44;
see \citep{zaccarella2015merge,friederici2017language}). The resulting assembly
would then represent the creation of the phrase ``hit the truck'', and would
have strong synaptic links to and from the two corresponding assemblies in
variable space.

We have presented a model which shows how binding capabilities emerge in
generic spiking neural networks through assembly projections. The model is
consistent with recent experimental data on assembly representations
\citep{quiroga2016neuronal} in cortex and the representation of thematic roles
in lmSTC \citep{FranklandGreene:15}.  Assembly projections can reconcile
functional needs, such as the recall of concepts, with data on the inherently
sparse connectivity between brain areas \citep{wang2016brain} and sparse
network activity.  The comprehensive repertoire of operations on assemblies of
neurons identified in the present paper, and the apparent effectiveness of
these operations, seem to give credence to an emerging hypothesis that
assemblies and their operations may underlie and enable many of the higher
mental faculties of the human brain, such as language, planning, story-telling,
reasoning, and science.

\section*{Acknowledgments}

This work was supported by the European Union project \#785907 (Human Brain
Project) and the Austrian Science Fund (FWF): I 3251-N33. The Titan Xp used for
this research was donated by the NVIDIA Corporation. We
thank Adam Marblestone for helpful comments.

{
\small
\putbib[549]
}
\end{bibunit}

\clearpage

\begin{bibunit}
\setcounter{section}{0}
\setcounter{table}{0}
\renewcommand\thesection{S\arabic{section}}
\renewcommand\thetable{S\arabic{table}}

\title{Supplementary Material for \textit{\titleref}}
\author{\authorshort}

\maketitle

\section{Details of neural network model}
\label{sec:network}

\paragraph{General network architecture}

The network consists of one content space $\C$ as well as one or several
variable spaces $\V, \U, \dots$ for variables $v, u, \dots$. Each neural space
consists of a pool of excitatory and a pool of inhibitory neurons (ratio: 4:1;
the number of excitatory neurons is $N = 1000$ for the content space and $N =
2000$ for all variable spaces in our simulations). Excitatory and inhibitory
neurons are sparsely interconnected (see below). Within each neural space, excitatory neurons are
connected by sparse recurrent connections with $p = 0.1$ (i.e., for each pair
of neurons, a connection between them is established with probability $p$,
independently of all other pairs). Excitatory neurons in each variable space
receive sparse excitatory connections ($p = 0.1$) from the excitatory neurons
in the content space $\C$ and vice versa. Since the connections are drawn at
random, they are generally asymmetric. Neurons in $\C$ additionally receive
input from an input population ($N_{\text{in}} = 200$ in our simulations).

\paragraph{Connections between excitatory and inhibitory neurons:}

In the following, $\mathcal{E}$ and $\mathcal{I}$ denote the pool of excitatory
and inhibitory neurons within a neural space, respectively. The parameters for
these connections are based on data from mouse cortex
\citep{avermann2012microcircuits}; the weights are
static and were calculated according to \citep{legenstein2017probabilistic} and
are given below in Tab.~\ref{tab:ei_params}.

\begin{table}[htbp]
\begin{center}
\begin{tabular}
{@{}lrrr@{}}
\toprule
connection & probability & synaptic weight & synaptic delay \\
\midrule
& & pA & ms \\
\midrule
  $\mathcal{E}\rightarrow\mathcal{I}$ & 0.575 &  17.39 & 0.5 \\
  $\mathcal{I}\rightarrow\mathcal{E}$ & 0.6   & -4.76  & 0.5 \\
  $\mathcal{I}\rightarrow\mathcal{I}$ & 0.55  & -16.67 & 0.5 \\
\bottomrule
\end{tabular}
  \caption{Connection parameters for static connections within each neural space. Given are the parameters for connections between the excitatory population $\mathcal{E}$ and the inhibitory population $\mathcal{I}$ as well as for recurrent connections from the inhibitory pool to itself. Recurrent excitatory connections are plastic and described in Tab.~\ref{tab:plastic_params}.}
\label{tab:ei_params}
\end{center}
\end{table}

\paragraph{Neuron model}

We used a single neuron model for all neurons in our simulations. In this
model, the probability of a spike of neuron $i$ at time $t$ is determined by
its instantaneous firing rate

\begin{equation}
\rho_i(t) = c_1 \cdot V_i'(t) + c_2 \cdot (\mathrm{e}^{c_3 \cdot V_i'(t)} - 1) \; ,
\end{equation}

\noindent where $c_1$, $c_2$, and $c_3$ can be used to achieve linear or
exponential dynamics (we use $c_1 = 0$, $c_2 = 1000\,\mathrm{Hz}$, $c_3 = 1 /
\mathrm{mV}$ for excitatory neurons and $c_1 = 10\,\mathrm{Hz} / \mathrm{mV}$
and $c_2 = c_3 = 0$ for inhibitory neurons). $V_i'(t)$ is the effective
membrane potential, which is calculated as $V_i'(t) = V_i(t) +
b_{i,\text{sfa}}(t)$, where $b_{i,\text{sfa}}(t)$ is an adaptive bias which
increases by some quantity $q_\text{sfa}$ every time the neuron spikes
(otherwise it decays exponentially with time constant $\tau_\text{sfa}$).
$b_{i,\text{sfa}}(t)$ is also clipped at some value $\hat{b}_\text{sfa}$ (in
our simulations, $q_\text{sfa} = 0.02\,\mathrm{mV}$, $\tau_\text{sfa} =
5\,\mathrm{s}$ and $\hat{b}_\text{sfa} = 0.5\,\mathrm{mV}$ for excitatory
neurons in variable spaces; for all other neurons $q_\text{sfa} = 0$). The
membrane potential is calculated using

\begin{equation}
V_i(t) = \mathrm{e}^{-\Delta t / \tau_\mathrm{m}} V_i(t - \Delta t) + (1 - \mathrm{e}^{-\Delta t / \tau_\mathrm{m}}) R_\mathrm{m} \left(I_{i,\mathrm{syn}}(t) + I_{i,\mathrm{inh}}(t - \Delta t) + I_\mathrm{e}\right)
\end{equation}

\noindent where $\tau_\mathrm{m}$ is the membrane time constant, $R_\mathrm{m}$
is the membrane resistance, and $I_\mathrm{e}$ is a bias current
($\tau_\mathrm{m} = 10\,\mathrm{ms}$, $R_\mathrm{m} = 0.5\,\mathrm{M}\Omega$;
furthermore, $I_\mathrm{e} = 0.2\,\mathrm{nA}$ for excitatory and $I_\mathrm{e}
= 0$ for inhibitory neurons). The current $I_{i,\mathrm{syn}}(t)$ results from
incoming spikes and is calculated via $I_{i,\mathrm{syn}}(t) = \sum_j w_{ij} z_j(t)$
where $z_j(t)$ are incoming spikes, and $w_{ij}$ are weights assigned to the
specific connection. $I_{i,\mathrm{inh}}$ controls the disinhibition and is set
to $-4\,\mathrm{nA}$ for all neurons inside a neural space if it is inhibited,
and $0$ otherwise.

After a neuron has spiked, its membrane potential is reset to zero, and the
neuron enters a refractory period. The duration of the refractory period (in
ms) is randomly chosen per neuron at the start of the simulation from a
$\Gamma$ distribution ($k = 4$, $\mu = 3.5$).

\paragraph{Plastic connections}

A simple model for STDP is used in the model for connections between excitatory
neurons. Each pairing of a pre- and a postsynaptic spike with $\Delta t =
t_\text{post} - t_\text{pre}$ leads to a weight change of

\begin{equation}
\Delta w(\Delta t) = \begin{cases} \eta \left( \mathrm{e}^{-\left|\Delta t\right| / \tau_+} - A_-\right) & \mbox{if } \Delta t \ge 0 \\
\eta \alpha \left( \mathrm{e}^{-\left|\Delta t\right| / \tau_-} - A_-\right) & \mbox{if } \Delta t < 0\end{cases}
\end{equation}

\noindent where $\tau_+, \tau_- > 0$ are time constants determining the width
of the learning window, $A_-$ determines the negative offset, $\alpha$
determines the shape of the depression term in relation to the facilitation
term and $\eta$ is a learning rate. This rule is similar to the one proposed by
\citep{NesslerETAL:2010}, but without a weight dependency. The parameters for
all plastic connections are given in Tab.~\ref{tab:plastic_params}. Weights
were clipped between 0 and an upper bound dependent on the connection type (see
Tab.~\ref{tab:plastic_params}). As we assume that disinhibition enables
learning, the learning rates for all synapses within a neural space were set to
zero during inhibition periods.

\begin{table}[htbp] \begin{center} \begin{tabular} {@{}rcrcrcllcccrcc@{}}
\toprule \multirow{2}{*}{connection} & \phantom{a} & \multirow{2}{10mm}{prob-
ability} & \phantom{a} & \multirow{2}{10mm}{synaptic delay} & \phantom{a}  &
\multicolumn{2}{c}{synaptic weight} & \phantom{a} &
\multicolumn{5}{c}{plasticity parameters} \\ \cmidrule{7-8} \cmidrule{10-14} &&
&& && init & bounds && $\alpha$ & $\tau_+$ & $\tau_-$ & $A_-$ & $\eta$ \\
\midrule && && ms && pA & pA && & ms & ms & & \\ \midrule
content space \\ $\mathcal{X}\rightarrow\mathcal{E}$ && 1   && 1, 10 && 0, 0.8
& 0, 0.8   && 0   & 25  &     & 0.4   & 0.01 \\
$\mathcal{V}\rightarrow\mathcal{E}$ && 0.1 && 1, 10 && 0.19\hl, 0.39\hl & 0,
0.87\hl && 0\hl  & 20\hl &     & 0.47\hl & 0.008\hl \\
$\mathcal{E}\rightarrow\mathcal{E}$ && 0.1 && 1     && 0            & 0, 0.6
&& -1  & 25  & 40\hl & 0.5   & 0.0025 \\ variable space \\
$\mathcal{C}\rightarrow\mathcal{E}$ && 0.1 && 1, 10 && 0.48\hl, 0.86\hl & 0,
1.33\hl && 0\hl  & 21\hl &     & 0.28\hl & 0.004\hl \\
$\mathcal{E}\rightarrow\mathcal{E}$ && 0.1 && 1     && 0.44\hl, 0.87\hl & 0,
1.08\hl && -1\hl & 37\hl & 49\hl & 0.52\hl & 0.006\hl \\ \bottomrule
\end{tabular} \caption{Connection parameters for all plastic connections in the
model. The parameters are given for incoming connections to the excitatory
neurons ($\mathcal{E}$) within the content space from the input population
($\mathcal{X}$) and from variable spaces ($\V$) as well as for recurrent
connections within the excitatory pool in the content space. For variable
spaces, the parameters for incoming connections from the content space ($\C$)
and for recurrent excitatory connections are given. Synaptic delays and initial
weights are drawn from uniform distributions within the given bounds.
Highlighted parameters (\hl) were determined using an optimization procedure
(see text).}
\label{tab:plastic_params}
\end{center}
\end{table}

\paragraph{Simulation:} Network dynamics were simulated using NEST
\citep{gewaltig2007nest,eppler2008pynest} with a time step of $\Delta t = 0.1$
ms.

\section{Details of experiments}

\paragraph{Initial formation of assemblies in the content space}

First, the content space learned to represent five very simple patterns
presented by the $200$ input neurons. Each pattern consisted of $25$ active
input neurons that produced Poisson spike trains at $100$ Hz while other
neurons remained silent (firing rate $0.1$ Hz), and each input neuron was
active in at most one pattern. This initial learning phase consisted of $200$
pattern presentations, where each presentation lasted for $200$ ms followed by
$200$ ms of random activity of the input neurons (all firing at $12.5$ Hz to
have a roughly constant mean firing rate of all input neurons). After the
training phase, synaptic plasticity of connections between the input population
and the content space as well as for recurrent connections within the content
space was disabled.

After the training phase, each pattern was presented once to the content space
for $200$ ms and the neuronal responses were recorded to investigate the
emergence of assemblies. If a neuron fired with a rate of $>50$ Hz during the
second half of this period, it was classified to belong to the assembly of the
corresponding input pattern. This yields five assemblies in $\C$; two of these
are shown in Fig.~\ref{fig:illustration} (showing a subset of neurons of $\C$, all
large weights between neurons belonging to some assembly, i.e. $>90\%$ of the
maximum weight, are shown with the color reflecting the assembly identity).

We created ten instances of such content spaces (i.e. random parameters such as
synaptic delays and recurrent connections were re-drawn) which were separately
trained for the experiments detailed below.

\paragraph{Creation of assembly projections (CREATE operation)}

Depending on the experiment, one or two variable spaces were added ($\V$ and
$\U$ for variables $v$ and $u$, each consisting of $2000$ excitatory and $500$
inhibitory neurons). Fig.~\ref{fig:illustration}A shows the potential connections
within the variable space $\V$ as well as potential feedforward and feedback
connections (all existing connections denoted by thin gray lines). Stable
assemblies were induced in each variable space separately by presenting each
input pattern to the content space for $1000$ ms (leading to the activation of
the encoding assembly there, Fig.~\ref{fig:illustration}B) while one of the variable
spaces was disinhibited (Fig.~\ref{fig:illustration}C). This was performed for every
reported execution of an operation to obtain results which are independent of
the specific network wiring.

Fig.~\ref{fig:illustration}C shows emerging assemblies (measured as in the content
space) in the variable space $\V$ during the CREATE phase for two different
contents (all large recurrent, feedforward, and feedback connections involving
the assembly in $\V$, i.e. the active neurons, drawn in color; dashed lines
denote feedback connections, i.e. from $\V$ to $\C$).

In all following figures, assemblies and connectivity are plotted as in
Fig.~\ref{fig:illustration}.

\paragraph{Optimization of plasticity parameters}

23 model parameters controlling synaptic plasticity (see \ref{sec:network},
Tab. \ref{tab:plastic_params}) were optimized using a gradient-free
optimization technique (see \ref{sec:optimization} for details). All parameters were
constrained to lie in a biologically reasonable range. We used the RECALL
operation (see below) to assess the quality of parameters.  The cost function
penalized mismatches of neural activity in both neural spaces during the CREATE
and RECALL periods. Using the obtained parameters, the RECALL operation could
reliably be performed across a variety of network architectures using the
success criterion  (see \ref{sec:optimization} and {\em Results} in main text for details).

\paragraph{Recall of content space assemblies (RECALL operation)}

To test whether content can be retrieved from variable spaces reliably, we
first presented a pattern to the network for $200$ ms with one of the variable
spaces $\V$ or $\U$ disinhibited. This corresponds to a brief CREATE operation.
Note that because assemblies in the variable spaces were already created
previously (see above), the previously potentiated synapses were still strong.
Hence, the shorter presentation period was sufficient to activate the assembly
in the variable space. In the following, we refer to such a brief CREATE as a
loading operation. After this loading phase, a delay period of $5$ s followed
(no input presented, i.e. input neurons firing at $12.5$ Hz). In order to make
sure that no memory was kept in the recurrent activity, all neural spaces were
inhibited in this period. After the delay, we retrieved the content of the
variable space in a RECALL operation, in which the variable space $\V$ was
disinhibited for $200$ ms.  During the first $50$ ms of this phase, the content
space remained inhibited.  Neurons in the content space were classified as
belonging to the assembly corresponding to the input pattern depending if their
firing rate was $>50$ Hz in the second half of the RECALL phase. A recall
operation was regarded as a success if $80$\% of the neurons in the
content space which belong to the assembly measured after training (see above)
were active during the RECALL phase while at the same time the number of excess
neurons (active during RECALL but not after training) did not exceed $20$\% of
the size of the original assembly.

\paragraph{Details to the binding of words to roles}

These experiments modeled the findings in \citep{FranklandGreene:15} about the
binding of content to roles in temporal cortex. We again used the network
described above with one content space $\C$ and two variable spaces, which we
refer to in the following as $\V_\text{agent}$ and $\V_\text{patient}$. Input
patterns to the network were interpreted as words in sentences. We used the
network described above with five assemblies in $\C$ that represented five
items (words) $A_1, \dots ,A_5$ and five assembly projections in each variable space
(created as described above). We defined that $A_1$ represents ``truck'' and
$A_2$ represents ``ball''. We considered the four binding sequences, corresponding to four sentences: 
S1=$\langle$ agent=''truck'', patient=''ball''$\rangle$, 
S2=$\langle$ patient=''ball'', agent=''truck''$\rangle$, 
S3=$\langle$ patient=''truck'', agent=''ball''$\rangle$, 
S4=$\langle$ agent=''ball'', patient=''truck''$\rangle$.
The processing of a sentence
was modeled as follows. The words ``truck'' and ``ball'' were presented to the
network (i.e., the corresponding input patterns) in the order of appearance in
the sentence, each for $200$ ms, without a pause in between. During the
presentation of a word, the activated assembly in $\C$ was bound to
$\V_\text{agent}$ if it served the role of the agent and to $\V_\text{patient}$
if its role was the patient. For example, for the sentence ``The truck hit the
ball'', first ``truck'' was presented and bound to $\V_\text{agent}$ (the
``agent'' variable), then ``ball'' was presented and bound to
$\V_\text{patient}$ (the ``patient'' variable).  The sequence of sentences S1
to S4 was presented twice to the network. The classifier described in the
following was trained on the first sequence and tested on the second sequence.

Spiking activity was recorded in all neural spaces. The first $50$ ms of each
word display were discarded to allow the activity to settle. Spiking activity
was then low-pass filtered with a filter time constant of $20$ ms to obtain the
filtered activity $r_i(t)$ for each neuron $i$ (as above, Eq.~\ref{eq:rit}).
Time was discretized with $\Delta t=1$ ms. Independent Gaussian noise (mean: 0,
variance: 4) was added at each time step to the trace of each neuron. We denote
by $\br_\text{agent}(t)$, $\br_\text{patient}(t)$, $\br_\mathcal{V}(t)$, and
$\br_\mathcal{C}(t)$ the vector of filtered activities with noise at time $t$
from all neurons in variable space $\V_\text{agent}$, in variable space
$\V_\text{patient}$, in both variable spaces, and in content space,
respectively.  The traces of neurons which never fired were discarded from
these vectors.

The task for the first classifier (``role of truck'', i.e. ``sentence decoder''
in Fig.~\ref{fig:fg}) was to classify the meaning of the current sentence at
each time point $t$ (this is equivalent to determining the role of the truck).
Hence, the sentences S1 and S2 constituted class $C_0$ and sentences S3 and S4
the class $C_1$. The classification was based on the current filtered network
activity $\br_\mathcal{V}(t)$ from the variable spaces. Using Scikit-learn
(version 0.19; \citealp{pedregosa2011scikit}), we trained a classifier for
logistic regression using the traces from the variable spaces. For comparison,
a classifier was also trained in the same manner on filtered network activity
$\br_\mathcal{C}(t)$ from the content space.

To model the second experiment from \citep{FranklandGreene:15}, we considered
sentences that were formed by tuples from the set of all five items $A_1,
\dots, A_5$, see {\em Results}.  Then, the task for a second classifier (``who
is the agent'') was use the filtered network activity ${\br}_\text{agent}(t)$
to classify the identity of the current agent during those times when
$\V_\text{agent}$ was disinhibited. The activity traces were created as in the
previous experiment. The data set was divided into a training set and a test
set as described in {\em Results}, and a classifier was trained (as above,
``agent decoder'' in Fig.~\ref{fig:fg}).  Finally, the task for a third
classifier (``patient decoder'' in Fig.~\ref{fig:fg}) was
to classify from subsampled filtered network activity ${\br}_\text{patient}(t)$
the identity of the current patient during those times when $\V_\text{patient}$
was disinhibited. The procedure was analogous to the procedure for the second
classifier.

\paragraph{Copying of assembly projections (COPY operation)}

After a content was loaded into $\V$ and a brief delay period ($400$ ms), a
RECALL operation was performed from $\V$ (duration $200$ ms as above). Then,
$\U$ was additionally disinhibited for $100$ ms. To test the performance, a
RECALL was initiated from $\U$ $400$ ms later (same success criterion as
above). We report the results on the same ten network instances as before.

\paragraph{Comparison of assembly projections (COMPARE operation)}

To test COMPARE operations in the circuit, a readout assembly consisting of
$50$ integrate-and-fire neurons was added to the network (resting potential
$-60$ mV, firing threshold $-20$ mV, membrane time constant $20$ ms, refractory
period of $5$ ms with reset of membrane potential) with sparse incoming
connections from excitatory neurons in the content space (connection
probability $0.1$) with depressing synapses (\citealp{markram1998differential};
parameters taken from \citealp{gupta2000organizing}, type F2; connection weight
$50$ pA).  After two load operations (duration $200$ ms, each followed by $50$
ms of random input) which store two assembly projections (to identical or
different contents in $\C$) in the variable spaces $\V$ and $\U$, we performed
a recall operation from each ($400$ ms in total, no delay). During this time,
the spike trains $S_i(t)$ for neuron i from the readout assembly were recorded
and filtered according to

\begin{equation}
\label{eq:rit}
r_i(t) = \int_0^{T_\mathrm{LP}} \mathrm{e}^{-\frac{s}{\tau_\mathrm{LP}}} S_i(t-s) ds
\end{equation}

\noindent to obtain the low-pass filtered activity for each neuron $i$
($T_\mathrm{LP} = 100$ ms, $\tau_\mathrm{LP} =$ 20 ms). The readout population
activity shown in Fig.~\ref{fig:compare}B,C was then calculated as
$R_\mathrm{readout}(t) = \sum_i r_i(t)$. There, all 25 possible comparisons
(between the five patterns available in $\C$) are shown.

\section{Details of parameter optimization}
\label{sec:optimization}

23 model parameters controlling synaptic plasticity (see
Tab.~\ref{tab:plastic_params}) were optimized using a gradient-free
optimization technique. All parameters were constrained to lie in a
biologically reasonable range.

We used the RECALL operation (see {\em Methods}) to assess the quality of
parameters.  The cost function penalized mismatches of neural activity in both
neural spaces during the CREATE and RECALL periods. We defined the set
$\mathcal{A}^\C_\mathrm{CREATE} = \{n_0, n_1, ...\}$ containing the neurons
$n_1, n_2, ...$ which were active in the content space $\C$ during the CREATE
operation. Similarily, we defined $\mathcal{A}^\C_\mathrm{RECALL}$ for the
neurons active in $\C$ during the subsequent RECALL, as well as
$\mathcal{A}^\V_\mathrm{CREATE}$ and $\mathcal{A}^\V_\mathrm{RECALL}$ for the
variable space. The cost in some iteration was then given by

\begin{gather}
C = \left|\mathcal{A}^C_\mathrm{CREATE}\bigtriangleup\mathcal{A}^C_\mathrm{RECALL}\right| + \lambda \left|\mathcal{A}^V_\mathrm{CREATE}\bigtriangleup\mathcal{A}^V_\mathrm{RECALL}\right|
\end{gather}

\noindent where $A \bigtriangleup B = (A \setminus B) \cup (B \setminus A)$ is
the symmetric difference between sets $A$ and $B$ and $\lambda = 10^{-4}$ is a
trade-off parameter.

The optimization procedure consisted of two steps: the initial parameter set
was obtained by evaluating 230 candidates from a Latin Hypercube sampler
\citep{mckay1979comparison} and choosing the parameter set with the lowest cost.
Then, a gradient-free optimization technique using a fixed number of epochs
($N_\mathrm{iter} = 500$) was used to further tune the parameters: in each
iteration, a subset of parameters was chosen for modification (Bernoulli
probability $p = 0.5$ per parameter). For each selected parameter, a new value
was drawn from a uniform distribution centered around the current value; the
width of these proposal distributions was decayed linearly from half of the
allowed parameter range (defined by the constraints as the difference between
upper and lower bound for each parameter value) in the first iteration to
$0.1\%$ of the allowed range in the last. (The proposal distributions were also
clipped to lie within the allowed range.) After evaluating the cost, the new
parameter set was kept if the cost was lower than the cost of the previous
parameter set, otherwise, the new parameters were discarded.

This optimization was performed using five pre-trained content space
instances: two were used for evaluating the costs and updating the parameters
(cost terms were generated in two separate runs and summed to get the overall
cost), three for early stopping (i.e. after the optimization, the parameter set
giving the lowest cost when tested on these three content spaces was chosen as
final parameter set). Using these parameters, the RECALL operation could
reliably be performed on the five content space instances used during the
optimization as well as on five new instances which were not used for optimiztion (i.e.
successful RECALL for each of the five values in each of the ten content space
instances, see below for the success criterion and  {\em Results} for details).

{
\small
\putbib[549]
}
\end{bibunit}

\end{document}